\documentstyle[preprint,aps]{revtex}
\begin{document}
\draft
\title{An extension of the coupled-cluster method: 
A variational formalism}
\author{Y. Xian}
\address{Department of Physics, UMIST\\
(University of Manchester Institute of Science and Technology)\\
P.O. Box 88, Manchester M60 1QD, UK}
\date{\today }
\maketitle

\begin{abstract}
A general quantum many-body theory in configuration space is developed
by extending the traditional coupled-cluster method (CCM) to a
variational formalism. Two independent sets (destruction and creation
sets) of distribution functions are introduced to evaluate the
Hamiltonian expectation. An algebraic technique for calculating these
distribution functions via two self-consistent sets of equations is
given. By comparing with the traditional CCM and with Arponen's
extension, it is shown that the former is equivalent to a {\it linear}
approximation to one set of distribution functions and the later is
equivalent to a (generalized) random-phase approximation to it. In
additional to these two approximations, other higher-order
approximation schemes within the new formalism are also discussed. As
a demonstration, we apply this technique to a quantum
antiferromagnetic spin model.
\end{abstract}


\section{introduction}

The main task of a microscopic quantum many-body theory is to study
correlations between the constituent particles of a quantum system in
a systematic way. The treatment of these many-body correlations is
either in real space or in configuration space. A real-space theory
usually focuses on the potential part of many-body Hamiltonians; a
configuration space theory often starts from the kinetic part of
Hamiltonians. One of the most successful real space quantum many-body
theories is the method of correlated basis functions (CBF) \cite{fee}
in which real-space correlation functions of the ground state are
determined variationally. Perhaps, the closest counterpart
of configuration space theories to the real space CBF is the
coupled-cluster method (CCM) \cite{hhc,ck,ciz} in which correlation
operators are employed to construct the ground state. One key feature
of the CCM is that the bra and ket states are not manifestly hermitian
to one another \cite{arp}.

In this paper we propose a general variational theory in configuration
space by extending the traditional CCM to a variational formalism in
which ket and bra states are hermitian to one another. The difficult
task of evaluating the Hamiltonian expectation can be done by
introducing distribution functions which can then be determined either
by a diagrammatic technique or by an algebraic one. The diagrammatic
approach developed in this context is quite similar to that of the
CBF. In the algebraic approach, one derives two similar sets of
self-consistent equations for the distribution functions; these
equations can then be tackled by various methods, e.g.  iterative
method. Easy comparison can be made with the traditional CCM in this
approach. We will mainly discuss the algebraic approach in this
article; the diagrammatic approach will be discussed elsewhere
\cite{xy}. We apply this variational method to a well-known spin model
as a demonstration. Some of our preliminary results has been reported
in a conference paper \cite{xy2}.

\section{The representation of a many-body wave function}

Similar to the method of correlated basis functions (CBF), the
coupled-cluster method (CCM) deals directly with the wave functions of
a many-body system. We shall take the spin-1/2 antiferromagnetic XXZ
model on a bipartite lattice as an example. The model Hamiltonian is
given by
\begin{equation}
H=\frac 12\sum_{l,\rho }H_{l,l+\rho}=
\frac 12\sum_{l,\rho }\left( \Delta s_l^zs_{l+\rho }^z+\frac 12%
s_l^{+}s_{l+\rho }^{-}+\frac 12s_l^{-}s_{l+\rho }^{+}\right) ,
\end{equation}
where $\Delta $ is the anisotropy, the index $l$ runs over all lattice
sites, $\rho $ runs over all nearest-neighbor sites, and $s^\pm$ are
the usual spin raising $(+)$ and lowering $(-)$ operators. The
Hamiltonian of Eq. (1)  at $\Delta =1$ corresponds to the isotropic Heisenberg
model which has been a focus of theoretical study in recent years due
to its relevance to  high-temperature superconductivity.

In the limit $\Delta \rightarrow \infty $, the ground state of Eq.~(1)
is clearly given by the classical N\'eel state with alternating
spin-up and spin-down sublattices. We shall exclusively use index $i$
for the spin-up sublattice and the index $j$ for the spin-down
sublattice. For a finite value of $\Delta$, such as the isotropic
point $\Delta =1$, the many-spin correlations in its ground state can
then be included by considering the excited states with respect to the
uncorrelated N\'eel model state. These excited states are constructed
by applying the so-called configuration creation operators $C^\dag_I$
to the N\'eel model state with the nominal index $I$ labelling these
operators. In our spin model, the operators $C^\dag_I$ are given by
any combination of the spin-flip operators to the N\'eel state, namely
$s^-_i$ and $s^+_j$ and the index $I$ in this case corresponds to the
collection of the lattice indices ($i$'s and $j$'s). The hermitian
conjugate operators of $C^\dag_I$ are the configuration destruction
operator $C_I$, given by any combination of $s^+_i$ and $s^-_j$. For
example, the two-spin flip creation operator is given by $C^\dag_{ij}
= s_i^- s_j^+$, and their destruction counterpart, $C_{ij}=s_i^+s_j^-$.

The traditional CCM is based on the Hubbard, Hugenholtz and Coester
representation (HHC) \cite{hhc} for the ground ket state, where the
correlations are parametrized by an exponentiated operator as,
\begin{equation}
\bigl| \Psi _g\bigr\rangle =e^S\bigl| \Phi \bigr\rangle, \quad
         S=\sum_I F_IC^\dag_I.
\end{equation}
For our spin model, $\bigl| \Phi \bigr\rangle $ is the N\'eel state and $F_I$ are
the correlation coefficients. The configuration creation operator $C^\dag_I$
in this case is given by a product of any number of pairs of the spin-flip
operators,
\begin{equation}
\sum_I F_I C^\dag_I= \sum_{n=1}^{N/2}
\sum_{i_1...,j_1...}f_{i_1...,j_1...}
\frac{s_{i_1}^{-}...s_{i_n}^{-}s_{j_1}^{+}...s_{j_n}^{+}}{(2s)^n},
\end{equation}
where $s$ is the spin quantum number. Although we are mainly
interested in $s=1/2$, we keep the factor of $1/2s$ for the purpose of
comparison with the large-$s$ expansion. Notice also that in Eq.~(3)
the spin-flip operators of the $i$-sublattice always pair with that of
the $j$-sublattice to ensure the total $z$-component $s^z_{total}=0$.
For the bra state, however, the CCM proposes a different, practical
form as \cite{ck,ciz,arp},
\begin{equation}
\bigl\langle\tilde\Psi\bigr| = 
\bigl\langle\Phi\bigr|\tilde S' e^{-S},
\end{equation}
where $S$ is as given in the ket state and the {\it linear} bra state
operator $\tilde S'$ is constructed by the configuration destruction
operators only, namely,
\begin{equation}
\tilde S' =1+\sum_I \tilde F_I\ C_I=1+\sum_{n=1}^{N/2}
\sum_{i_1...,j_1...}\tilde f_{i_1...,j_1...}
\frac{s_{i_1}^{+}...s_{i_n}^{+}s_{j_1}^{-}...s_{j_n}^{-}}{(2s)^n}\;.
\end{equation}
The coefficients $\{F_I,\tilde F_I\}=\{f_{i_1...j_1...}, \tilde
f_{i_1...j_1...}\}$ are determined variationally through the
Hamiltonian expectation $\bigl\langle H\bigr\rangle$, noticing the 
normalization condition $\bigl\langle\tilde\Psi |
\Psi\bigr\rangle=1$,
\begin{equation}
\bigl\langle\tilde\Psi\bigr| H\bigl|\Psi\bigr\rangle
=\bigl\langle\Phi\bigr|\bar H\bigl|\Phi\bigr\rangle,
\end{equation}
where the similarity-transformed $\bar H=e^{-S}He^S$ can be expanded as a series of
nested commutators as
\begin{equation}
\bar H = H+\frac{1}{1!}[H,S]+\frac{1}{2!}[[H,S],S]+\cdots\;.
\end{equation}
In most cases, $H$ contains a finite order of destruction operators.
The above series then terminates at a finite order
as $S$ contains only the creation operators. Hence, the Hamiltonian
expectation value in the CCM is a finite order polynomial function of
the coefficients $\{F_I,\tilde F_I\}$. More specifically,
$\bigl\langle H\bigr\rangle$ in the CCM is linear in the bra-state
coefficients  $\tilde F_I$ and finite-order polynomial in the
ket-state coefficients  $F_I$. Thus, calculations in the CCM in
general are quite straightforward; when an approximation scheme is
chosen (i.e., a truncation scheme with a finite set of 
$\{F_I,\tilde F_I\}$), no further
approximation is necessary in most calculations. However, this CCM
parametrization of the ground state is problematic in dealing with
long-range correlations as discussed in the context of our spin model
calculations in Ref.~\cite{bpx}. More discussion of the problems in 
this traditional CCM will be given later.

An obvious extension of the CCM is to apply the HHC representation to both
the ket and bra states. Hence we write
\begin{eqnarray}
\bigl| \Psi \bigr\rangle &=& e^S\bigl| \Phi \bigr\rangle,\quad
        S=\sum_I F_I C^\dag_I, \\
\bigl\langle \tilde\Psi \bigr| &=&\bigl\langle \Phi \bigr|
e^{\tilde S},\quad
        \tilde S=\sum_I \tilde F_I C_I,
\end{eqnarray}
where the ket-state and bra-state correlation coefficients $F_I$ and
$\tilde F_I$ are hermitian and independent to one another. For our
spin model, the model state $\bigl|\Phi\bigr\rangle$ is the N\'eel
state, and the correlation operators $\sum_I F_IC^\dag_I$ and $\sum_I
\tilde F_IC_I$ are given as in Eqs.~(3) and (5). The coefficients
$\{F_I,\tilde F_I\}$ are then determined by the usual variational
equations as
\begin{equation}
\frac{\delta\bigl\langle H\bigr\rangle}{\delta \tilde F_I} =
\frac{\delta\bigl\langle H\bigr\rangle}{\delta F_I} = 0\;,
\end{equation}
where energy expectation is defined in the usual way as
\begin{equation}
\bigl\langle H\bigr\rangle = 
\frac{ \bigl\langle\tilde\Psi\bigr| H \bigl|\Psi\bigr\rangle}
{\bigl\langle\tilde\Psi | \Psi\bigr\rangle} =
\frac{ \bigl\langle\Phi\bigr|e^{\tilde S} H e^S\bigl|\Phi\bigr\rangle}
{\bigl\langle\Phi\bigr|e^{\tilde S}e^S\bigl|\Phi\bigr\rangle}\;.
\end{equation}
Clearly, the normalization factor $\bigl\langle\tilde\Psi|
\Psi\bigr\rangle$ and Hamiltonian expectation $\bigl\langle
H\bigr\rangle$ are highly nontrivial functions of the coefficients
$F_I$ and $\tilde F_I$. Their calculation in the standard variational
approach is in general difficult, contrast to the CCM where the
expectation value of the Hamiltonian is a finite-order polynomial of
the coefficients, as described earlier. This is perhaps the main
reason that little progress has been made in this extension of the CCM
except some general discussion and a few attempts \cite{kutz}.

Hence, the key to the extension of the CCM to the standard variational
method as described in Eqs.~(8)-(11) is to develop a practical and
consistent technique to evaluate the normalization factor and the
Hamiltonian expectation. It is known in statistical mechanics and
real space quantum many-body theory that these evaluations can be
done more efficiently by employing distribution functions. One then
needs to develop a systematic and consistent scheme to calculate these
distribution functions. We have considered two such schemes. One is
similar to the traditional technique in statistical mechanics employed
by the CBF. In this method, one introduces a generating functional
whose functional derivatives are the distribution functions. A
diagrammatic technique has been developed to evaluate these
distribution functions \cite{xy}.

The other approach we have considered is an algebraic technique. In
this approach, one derives two similar self-consistent sets of
equations for the (destruction and creation) distribution functions by
taking the advantage of the operator nature in the ground state as
given in Eqs.~(8)-(9). These self-consistent set of equations can then
be tackled by various methods such as iteration method.  As we shall
see, a most simple approximation to one of these two self-consistent
sets of equations reproduces the full CCM results, but one can easily
go beyond that.  We shall mainly discuss the algebraic approach in the
followings.

\section{Distribution functions and their self-consistent equations}

We first introduce the so-called bare distribution functions as
expectation value of the configuration operators, namely
\begin{equation}
g_I=\bigl\langle C^\dag_I\bigr\rangle,\quad
\tilde g_I=\bigl\langle C_I \bigr\rangle\;,
\end{equation}
where the expectation value is defined in the usual sense as in
Eq. (11). In general these bare distribution functions are nontrivial
functions of $\{F_J,\tilde F_J\}$. Multiplying by the corresponding
coefficients, we obtain $F_I g_I$ and $\tilde F_I \tilde g_I$ which
are the the usual full distribution functions useful in the
diagrammatic approach\cite{xy}.

Direct calculation of these functions is certainly not an easy task.
Fortunately, by taking the advantage of the properties of the operators, one
can derive self-consistent sets of equations which can then be
tackled by various methods. In particular, as $C^\dag_I$
commutes with $S=\sum_I F_IC^\dag_I$ and $C_I$ with $\tilde S= \sum_I
\tilde F_I C_I$, one can write
\begin{eqnarray}
g_I&=&\frac{1}{A} \bigl\langle\Phi\bigr| e^{\tilde S} C^\dag_I
    e^S\bigl|\Phi\bigr\rangle
   =\frac{1}{A} \bigl\langle\Phi\bigr| e^{\tilde S}e^S
    C^\dag_I\bigl|\Phi\bigr\rangle\;, \\
\tilde g_I&=&\frac{1}{A} \bigl\langle\Phi\bigr| e^{\tilde S} C_I
    e^S\bigl|\Phi\bigr\rangle
   =\frac{1}{A} \bigl\langle\Phi\bigr| C_I e^{\tilde S}e^S
    \bigl|\Phi\bigr\rangle\;,
\end{eqnarray}
where $A=\bigl\langle\tilde\Psi | \Psi\bigr\rangle$ is the
normalization constant.  In order to find another expression, one
inserts the identity $e^{-\tilde S}e^{\tilde S}$ in the expression of
$g_I$ as
\begin{equation}
g_I=\frac{1}{A} \bigl\langle\Phi\bigr| e^{\tilde S} C^\dag_I
    e^{-\tilde S}e^{\tilde S}e^S\bigl|\Phi\bigr\rangle
   =\frac{1}{A} \bigl\langle\Phi\bigr|\underbar C^\dag_I
     e^{\tilde S}e^S\bigl|\Phi\bigr\rangle\;,
\end{equation}
where the similarity-transformed operator $\underbar C^\dag_I$
can be expanded in the nested commutator series as
\begin{equation}
\underbar C^\dag_I = e^{\tilde S}C^\dag_I e^{-\tilde S}
 = C^\dag_I + \frac{1}{1!}[\tilde S, C^\dag_I]
    +\frac{1}{2!}[\tilde S,[\tilde S, C^\dag_I] + \cdots\;,
\end{equation}
and this series is finite as $C^\dag_I$ is finite and $\tilde S$
contains only the destruction operators. By definition,
$\bigl\langle\Phi\bigr|C^\dag_I = 0$, hence
$\bigl\langle\Phi\bigr|\underbar C^\dag_I$ can be expressed in a form
linear in the destruction operators $C_J$ and finite order polynomial
in the coefficients $\tilde F_J$. The expectation values of these
finite order terms is therefore linear in $\tilde g_J$ and
finite order polynomial in $\tilde F_J$. This yields a linear
relation between $g_I$ and $\{\tilde g_J\}$. Hence we write
\begin{equation}
g_I=G\bigl(\{\tilde g_J\},\{\tilde F_J\}\bigr)\;,
\end{equation}
where $G$ is a function linear in $\tilde g_J$ and finite order
polynomial in $\tilde F_J$.

In a similar fashion we write,  by inserting identity $e^Se^{-S}$ in the
expression for $\tilde g_I$,
\begin{equation}
\tilde g_I = \frac{1}{A}\bigl\langle\Phi\bigr|
            e^{\tilde S}e^S\bar C_J\bigl|\Phi\bigr\rangle\;,
\end{equation}
with the usual commutation series
\begin{equation}
\bar C_I = e^{-S}C_Ie^S = C_I+\frac{1}{1!}[C_I,S]
            +\frac{1}{2!}[C_I,S],S] + \cdots\;,
\end{equation}
and we obtain
\begin{equation}
\tilde g_I= G\bigl(\{g_J\},\{F_J\}\bigr)\;,
\end{equation}
where $G$ is the same function as in Eq.~(17) but now linear in
$\{g_J\}$ and finite order polynomial in $\{F_J\}$. As function $G$ is
the same in Eqs.~(17) and (20), only one calculation is necessary.

Eqs.~(17) and (20) provide two self-consistent sets of equations for $g_I$
and $\tilde g_I$ in terms of the correlation coefficients $\{F_J,
\tilde F_J\}$. We note that for a particular $g_I$ its equation in
general contains a higher-order set $\{\tilde g_J\}$, and vice versa,
even for a truncated coefficient set $\{F_I,\tilde F_I\}$. Therefore,
in order to make any practical calculation, one has to make two
approximations, a truncation on the number of coefficients
$\{F_I,\tilde F_I\}$, and a truncation on the number of bare
correlation functions $\{g_I,\tilde g_I\}$. After these two
truncations, one should be able to solve the self-consistent set of
equations to obtain $g_I$ and $\tilde g_I$ in terms of $F_I$ and
$\tilde F_I$. This is contrary to the CCM where one needs only one
truncation (in $\{F_I,\tilde F_I\}$). As we shall see, for a similar
truncation in the coefficients $\{F_I,\tilde F_I\}$, one of our lowest
order truncations in the bare correlation function $\{g_I,\tilde
g_I\}$ will reproduce the full CCM results. However, it is a simple step
to go beyond this approximation by including some higher-order
distribution functions $g_I$ which has proved to be essential to
obtain the consistent long-range behaviors of the spin correlation
functions (and the low-lying excitation energies) as we shall see in
our spin model calculation.

Since Hamiltonian usually contains terms involving both creation
operators $C_I^\dag$ and destruction operators $C_I$, using the
similar argument to the distribution discussed above, it is not
difficult to obtain that the expectation value of a general
Hamiltonian can be expressed as a function linear in $g_I$ and $\tilde
g_I$ and a finite order polynomial in $F_I$ or $\tilde F_I$,
\begin{equation}
\bigl\langle H\bigr\rangle =
          {\cal H}\bigl(\{g_I\},\{\tilde g_I\},\{F_I\}\bigr)=
          {\cal H}\bigl(\{\tilde g_I\},\{g_I\},\{\tilde F_I\}\bigr)\;.
\end{equation}
This expression is not unique; using Eqs.~(17) or (20), one can
express the Hamiltonian expectation as a function linear in $g_I$ and
finite-order polynomial in $F_I$ only, or as a function linear in
$\tilde g_I$ and finite-order polynomial in $\tilde F_I$ only,
\begin{equation}
\bigl\langle H\bigr\rangle ={\cal H}'\bigl(\{g_I\},\{F_I\}\bigr)
                    ={\cal H}'\bigl(\{\tilde g_I\},\{\tilde F_I\}\bigr)\;.
\end{equation}
This expression of the Hamiltonian expectation is useful when we
compare with the traditional CCM. Solutions of Eqs.~(17) and (20) can
be substituted into these equations and we obtain $\bigl\langle
H\bigr\rangle$ as a function of $\{F_I, \tilde F_I\}$. Variational
calculation in Eq.~(10) can then be carried out. In the following, we
consider a simple application to the spin model as a demonstration.

\section{Two-Spin Flip Approximation in a Spin Model}

As a demonstration, we consider a simple truncation approximation in
which the correlation operators $S$ and $\tilde S$ retain only the
two-spin flip correlations as
\begin{equation}
\bigl| \Psi _2\bigr\rangle =e^{S_2}\bigl| \Phi 
            \bigr\rangle, \quad
\bigl\langle \Psi _2\bigr| =\bigl\langle \Phi \bigr| e^{\tilde S_2} \;,
\end{equation}
where
\begin{equation}
S_2=\sum_{ij}f_{ij}\frac{s_i^{-}s_j^{+}}{2s},\quad 
\tilde S_2=\sum_{ij}\tilde f_{ij}\frac{s_i^{+}s_j^{-}}{2s}\;. 
\end{equation}
Using the usual angular momentum commutations
\begin{equation}
\bigl[s^z_l,s^\pm_{l'}\bigr]=\pm s^\pm_l\delta_{ll'},\quad
\bigl[s^+_l,s^-_{l'}\bigr]=2s^z_l\delta_{ll'}\;,
\end{equation}
and the N\'eel state eigenequations,
$s_i^z\bigl|\Phi\bigr\rangle = s\bigl|\Phi\bigr\rangle,
s_j^z\bigl|\Phi\bigr\rangle = -s\bigl|\Phi\bigr\rangle$,
it is a straightforward calculation to 
derive expectation value of various operators with respect to the states
of Eqs.~(23). In this approximation, for example, the order parameter is 
derived as
\begin{equation}
M^z=\bigl\langle s^z_i\bigr\rangle = s-\sum_r n_r\;,
\end{equation}
where $n_r$ is the full one-body distribution function given by
\begin{equation}
n_{ij}=f_{ij}g_{ij}=f_{ij}\frac{\bigl\langle s^-_is^+_j\bigr\rangle}{2s}\;,
\end{equation}
and we have taken the advantage of translational invariance by writing
$n_{ij}=n_r$ with $j=i+r$; the usual two-spin correlation function is
given by
\begin{equation}
\bigl\langle s^z_is^z_j\bigr\rangle = -s^2 
    +s\bigl(\sum_{i'}n_{i'j}+\sum_{j'}n_{ij'}\bigr)
    -\bigl(\sum_{i'j'}G_{ij',i'j}+n_{ij}\bigr)\;,
\end{equation}
where $G_{ij,i'j'}$ is the full two-body distribution function
\begin{equation}
G_{ij,i'j'}=f_{ij}f_{i'j'}g_{ij,i'j'}=
   f_{ij}f_{i'j'}\frac{\bigl\langle s^-_is^+_js^-_{i'}s^+_{j'}
         \bigr\rangle}{(2s)^2}\;;
\end{equation}
and finally, the expectation value of Eq.~(1) is then given by
\begin{equation}
\bigl\langle H_{ij}\bigr\rangle= -\Delta s^2 
 +s\biggl(\tilde g_{ij}+g_{ij}+\Delta\sum_{i'}n_{i'j}
  +\Delta\sum_{j'}n_{ij'}\biggr)
 -\Delta \biggl(\sum_{i'j'}G_{ij',i'j}+n_{ij}\biggr)\;.
\end{equation}
As can be seen, these physical quantities involve up to two-body
distribution functions.

The self-consistent set of equations for the bare distribution functions
are derived as described in Sec.~III. In particular, the equation
for the one-body function $\tilde g_{ij}$ is
\begin{eqnarray}
\tilde g_{i_1j_1}
&=&f_{i_1j_1}+\sum_{ij}f_{ij_1}f_{i_1j}g_{ij} \nonumber\\
&&\ -\frac 2{2s}f_{i_1j_1}\bigl( \sum_if_{ij_1}g_{ij_1}
+\sum_jf_{i_1j}g_{i_1j}\bigr) \nonumber  \\
&&+\frac 1{2s}\sum_{ijj^{\prime }}f_{ij_1}f_{i_1j}f_{i_1j^{\prime }}
g_{i_1j,ij'}+\frac 1{2s}\sum_{ii^{\prime }j}f_{ij_1}f_{i^{\prime
}j_1}f_{i_1j}g_{ij_1,i'j} \nonumber\\
&&\ +\frac 2{\bigl( 2s\bigr) ^2}f_{i_1j_1}^2g_{i_1j_1}
+\frac 4{\bigl( 2s\bigr)
^2}f_{i_1j_1}\sum_{ij}f_{ij_1}f_{i_1j}g_{i_1j_1,ij} \nonumber \\
&&+\frac 1{\bigl( 2s\bigr) ^2}\sum_{ii^{\prime }jj^{\prime
}}f_{ij_1}f_{i^{\prime }j_1}f_{i_1j}f_{i_1j^{\prime }}g_{i_1j_1,ij,i'j'}\;.
\end{eqnarray}
The equation for the two-body function $\tilde g_{ij,i'j'}$ will
contain up to twelve-body functions, etc. The hermitian conjugate of
these equations are the self-consistent set of equations for $g_{ij},
g_{ij,i'j'}$, etc. Clearly, we need to make further truncations for
any practical calculation.

Consider a simple truncation in which we retain only the first two
terms in Eq.~(31), noticing that all other terms are higher-order
in terms of $1/2s$ expansion,
\begin{equation}
\tilde g_{i_1j_1}\approx f_{i_1j_1}+\sum_{ij}f_{ij_1}f_{i_1j}g_{ij}\;,
\end{equation}
and similar equation for $g_{ij}$. Using the Fourier transformation
technique and translational symmetry, it is easy to solve the two
equations to obtain
\begin{equation}
g_k=\frac{f_k}{1-f_k\tilde f_k},\quad 
\tilde g_k=\frac{\tilde f_k}{1-f_k\tilde f_k}\;,
\end{equation}
where $g_k$ and $f_k$ are Fourier transformations of $g_{ij}$ and
$f_{ij}$, etc. To the same order in $1/2s$, the ground-state energy,
Eq.~(30), is
\begin{equation}
e=\frac{2}{zN}\sum_{i,\rho}\bigl\langle H_{i,i+\rho}\bigr\rangle
=-\Delta s^2+s(g_1+\tilde g_1)+2\Delta s \sum_r n_r\;,
\end{equation}
where $z$ is the number of nearest-neighbor sites and
\begin{equation}
g_1=\sum_k\frac{\gamma_kf_k}{1-\tilde f_kf_k},\quad
\tilde g_1=\sum_k\frac{\gamma_k\tilde f_k}{1-\tilde f_kf_k},\quad
\sum_r n_r=\sum_k\frac{f_k\tilde f_k}{1-\tilde f_kf_k}\;,
\end{equation}
with
\begin{equation}
\gamma_k = \frac1z \sum_\rho e^{i{\bf k\cdot r_\rho}}\;.
\end{equation}
The variational equations, $\frac{\partial e}{\partial
f_k}=\frac{\partial e}{\partial\tilde f_k}=0$, reduce to a quadratic
equation for $f_k$ and $\tilde f_k$. The physical solution to these
equations is
\begin{equation}
f_k=\tilde f_k=\frac{\Delta}{\gamma_k}
\bigl(-1+\sqrt{1-\gamma_k^2/\Delta^2}\bigr)\;.
\end{equation}
The ground-state energy and order parameter are then obtained as
\begin{equation}
e=-\Delta s^2+s\Delta \sum_k \bigl(-1+\sqrt{1-\gamma_k^2/\Delta^2}\bigr)
\end{equation}
and
\begin{equation}
M^z=s-\frac12\sum_k\bigl(\frac{1}{\sqrt{1-\gamma_k^2/\Delta^2}}-1\bigr)\;.
\end{equation}

It is not difficult to include the contribution of higher-order
many-body distribution functions within our variational
formalism. Consider the two-spin correlation function of Eq.~(28),
which contain the important full two-body distribution function.  The
bare two-body distribution functions $g_{ij,i'j'}$ and $\tilde
g_{ij,i'j'}$ can be calculated through their self-consistent set of
equations by keeping the same order terms in the $(1/2s)$ expansion as
we have done for the one-body distribution function in
Eqs.~(31)-(32). Without going into details of derivation, to the same
approximation, we obtain the following results
\begin{equation}
g_{ij,i'j'}\approx g_{ij}g_{i'j'}+g_{ij'}g_{i'j}
\end{equation}
which, in fact, is usually referred as random-phase approximation.  A
simpler way to obtain the same results for the two-body functions is
to employ the following sequential equation
\begin{equation}
\frac{\partial}{\partial f_{i'j'}}g_{ij} =g_{ij,i'j'}-g_{ij}g_{i'j'}\;,
\end{equation}
to $g_{ij}$ and $\tilde g_{ij}$ in Eq.~(32) and its hermitian
conjugate.

Hence, the normalized two-spin correlation function becomes
\begin{equation}
c_r=\bigl\langle s^z_is^z_{i+r}\bigr\rangle-
     \bigl\langle s^z_i\bigr\rangle\bigl\langle s^z_{i+r}\bigr\rangle
   =-g_r\tilde g_r =-g_r^2\;.
\end{equation}
In fact, our above results of the ground-state energy, order parameter
and the correlation function are the same as that of the spin-wave
theory \cite{and,tak}. In particular, the long-range behavior of the
correlation function $c_r\propto 1/r^2$ as $r\rightarrow\infty$ for a
square lattice system at $\Delta=1$ can not be obtained without the
contribution of the two-body distribution function. Using iteration
method for solving the equations of $g_{ij}$ and $\tilde g_{ij}$, it
is straightforward to include higher-order contribution for other
physical quantities. But we refrain ourselves from more detailed
calculation in this article as our main purpose here is to introduce
the new variational formalism and its comparison with the traditional
CCM.

\section{Comparison with the CCM}

In order to make a more detailed comparison with the traditional CCM,
we first summarize our variational extension. We apply the HHC
representation of Eqs.~(8)-(9) for the ground state wave function of a
quantum many-body system to {\it both} the ket and bra states with two
independent, hermitian conjugate correlation coefficients
$\{F_I,\tilde F_I\}$ which are determined by the variational equations
as
\[
\frac{\partial\langle H\rangle}{\partial F_I} = 
\frac{\partial\langle H\rangle}{\partial\tilde F_I} =0\;.
\]
The difficult task of expressing the Hamiltonian expectation 
$\langle H\rangle = {\cal H}(\{F_I,\tilde F_I\})$ can be done by 
introducing the bare distribution functions 
$g_I=\langle C_I^\dagger\rangle$ and $\tilde g_I=\langle C_I\rangle$
and solving their self-consistency equations
\[
g_I=G(\{\tilde g_J\},\{\tilde F_J\}),\quad
\tilde g_I=G(\{g_J\},\{F_J\})\;.
\]

Using the expression of Eq. (22) for the Hamiltonian expectation
\[
\bigl\langle H\bigr\rangle ={\cal H}'\bigl(\{g_I\},\{F_I\}\bigr)\;,
\]
we see  that the traditional CCM is equivalent to the {\it linear
approximation} in the self-consistensy equation for $g_I$, namely
\begin{equation}
g_I\approx \tilde F_I\;,
\end{equation}
for all possible values of $I$; the Hamiltonian expectation is then
reduced to a simple form
\begin{equation}
\bigl\langle H\bigr\rangle \approx
       {\cal H}_{\rm CCM}\bigl(\{\tilde F_I\};\{F_I\}\bigr)\;,
\end{equation}
where function ${\cal H}_{\rm CCM}$ is linear in $\tilde F_I$ and finite
order in $F_I$. In our spin model calculation of Sec. IV, within the
similar truncation involving only up to two-spin flip correlations
(the so-called SUB2 approximation), the corresponding CCM calculation is
to ignore all two-body and higher-order many-body distribution
functions in Eq.~(31). The two-spin correlation function thus calculated
has unphysical behaviors as discussed in Ref. 8.

In Arponen's extension of the CCM \cite{arp}, while keeping the
ket state in the traditional CCM form, the bra state, also not manifestly
hermitian to ket state, is parametrized by a nonlinear factor as
\begin{equation}
\langle\tilde \Psi| = \langle \Psi|e^{\tilde S}e^{-S}\;,
\end{equation}
where correlation operators $S$ and $\tilde S$ are give as in
Eqs.~(8)-(9). Within our variational formalism, as can be
demonstrated, this representation of the bra state is equivalent to
applying the random-phase approximation to the $g_I$ equation to
obtain $g_I$ as finite-order polynormial of $\{\tilde f_J\}$,
\begin{equation}
g_I\approx g_I(\{\tilde f_J\})\;.
\end{equation}
A detailed calculation in our spin model revealed that in the similar
SUB2 truncation in Arponen's approach, the random-phase results of
Eq.~(40) for the two-body distribution function is reproduced. While
it is clearly an improvement over the traditional CCM, Arponen's
extension is also known to be poor for the strongly correlated systems
(e.g., quantum Helium-4 fluid)even when high-order many-body
contributions are considered. This may be related to the random-phase
approximation of Eq.~(46).

In conclusion, our variational extension of the CCM provides a general
many-body theory in which the traditional CCM represents a simple
linear approximation and Arponen's extension represents a random-phase
approximation. The traditional CCM is well known to be efficient in
obtaining accurate ground state energy for a finite system with a
large energy gap separating the ground and excited states \cite{bb};
but, as pointed out earlier, it is poor when long-range correlations
in the system are important; and Arponen's extension provides a remedy
in producing the physical long range behaviors. But both these two
methods are known to be poor in dealing with strongly correlated
systems which demand correct description at short-range, in most cases,
strongly repulsive. We believe the variational formalism represented
here may provide an effective approach to the strongly correlated
systems. Furthermore, its strong overlap with other well known
many-body theories such as the method of correlated basis functionals
can provide useful clues in making suitable approximations in
practical calculations. In this regard, we are encouraged to read a
recent preprint \cite{fan} in which a fully variational approach has
been employed to study several weakly interacting boson systems using
the so-called independent pair correlation functions, which, in fact,
is the corresponding SUB2 truncation of the Eq.~(23) but writing in
the real space. We like to point out that a clear advantage of the
algebraic approach presented here is that it is straightfowardly
extendible to include higher-order many-body correlations beyond the
SUB2 level and as well as to other systems as electrons.

\acknowledgments Many useful discussions with J. Arponen, R.F. Bishop,
F. Coester, and H. K\"ummel are acknowledged.

\end{document}